\documentclass{raa}            

\usepackage{natbib}

\newcommand{\aph}{    {\it Astropart. Phys.}}

\usepackage{multirow}
\usepackage{ulem}
\usepackage{graphicx,times}             

\begin{document}

   \title{A CME-driven shock analysis of the 14-Dec-2006 SEP event
}

   \volnopage{Vol.0 (200x) No.0, 000--000}      
   \setcounter{page}{1}          

   \author{Xin Wang
      \inst{1,2}
   \and Yihua Yan
      \inst{1}
   }

   \institute{National Astronomical Observatories, Chinese Academy of Sciences,
             Beijing 100012, China; {\it wangxin@nao.cas.cn}\\
        \and
            State Key Laboratory of Space Weather, Chinese
Academy of Sciences, Beijing 100080, China \\
{\it Received 2012 April 12; accepted 2012 May 7} }

\abstract{{Observations of the interplanetary shock provide us with
strong evidence of particle acceleration to multi-MeV energies, even
up to GeV energy, in a solar flare or coronal mass ejection (CME).
Diffusive shock acceleration is an efficient mechanism for particle
acceleration. For investigating the shock structure, the energy
injection and  energy spectrum of a CME-driven shock, we perform
dynamical Monte Carlo simulation of the 14-Dec-2006 CME-driven shock
using an anisotropic scattering law. The simulated results of the
shock fine structure, particle injection, and energy spectrum are
presented. We find that our simulation results give a good fit to
the observations from multiple spacecraft.}  }

   \authorrunning{X. Wang \& Y. Yan}            
   \titlerunning{SIMULATION OF A CME-DRIVEN SHOCK }  

   \maketitle

%
%
\section{Introduction}           
\label{sect:intro}It is widely accepted that there are two classes
of solar energetic particle (SEP) events, although recent
observations indicate that the actual processes may be much more
complicated \citep{PV2008}.  The first class is normal impulsive SEP
events, which are connected with the large solar flare
\citep{Miller1997}. The second class is gradual SEP events, which
are responsible by diffusive shock acceleration (DSA) associating
with fast coronal mass ejections (CMEs)\citep{CRR1991,YPW2006}.
 In solar magnetic connection region, CME and
flares are two type of manifestations of the same magnetic energy
release process \citep{WSWL1996,ZDH2001,ZSW2007}. Both CMEs and
flares result in particle acceleration that constitute an SEP event.
But which manifestation dominates the particle injection is still
not clear \citep{LDV2009,LCW2011,QS2009}. Some numerical models
suggest that mixed particle acceleration by both flares and
CME-driven shocks provide much better fits to the in-situ
observations. Since the particle injection process is connected with
the complicated nonlinear effects in the particle acceleration
processes and also there exists difference of injection mechanism in
two type of manifestations, so we just put forward to a pure shock
numerical model to calculate the particle injection in the
CME-driven shock. We expect that it would be helpful for
understanding the particle injection problem in the SEP events.

DSA theory was first introduced in the later of 1970's (
\cite{krymsky77,axford77,bell78,bo78}. In the past several decades,
the accumulation of the increasingly observational data from many
spacecraft investigated the nonlinear diffusive shock acceleration
(NLDSA) mechanism, which is the most efficient accelerator in many
astrophysical and space physical environments
\citep{bo99,md01,bub09,bt11,LXW2006,ZBH2006}. With the development
of technology in observational equipments, especially in spacecraft
working in deep space, there are a lot of models for modeling the
various nonlinear interaction with the diffusive shock acceleration.
Several main approaches for studying the nonlinear DSA includes: the
two-fluid model \citep{DV1981,DAS1982}; the numerical model
\citep{bv00,kj07,zirak07,vlz10}; the stationary or dynamical Monte
Carlo model \citep{ee84,kje96,veb06}; the semi-analytical model
\citep{mdv00,cab10b} and etc. Among these approaches, the Monte
Carlo method addresses the nonlinear effects of DSA by assuming that
the entire particle population undergoes a random walk under a
certain scattering law \citep{emp90,kje96,wy11}.

There are three important non-linear processes of DSA theory
including the particle injection, particle confinement, and shock
robustness \citep{md01,Hu2009}. Owing to the fact that walking
processes of the particles can be controlled self-consistently in
the Monte Carlo method, the Monte Carlo method has an advantage for
simulating the particle injection. We have already studied the
energy translation processes of an Earth-bow shock using the
dynamical Monte Carlo method with a prescribed multiple anisotropic
scattering angular distributions \citep{wy12}. We find that the
acceleration efficiency increases as the dispersion of the
scattering angular distribution increases from an anisotropic case
to an isotropic case. Here, we will further investigate this
important particle injection problem in the CME-driven shock using
the dynamical Monte Carlo method. There exist a few different
properties between the Earth-bow shock and the CME-driven shock:
Firstly, Earth-bow shock has a stationary downstream bulk flow  but
the CME-driven shock has a dynamical downstream bulk flow; Secondly,
the CME-driven shock front has an opposite motion compared with the
Earth-bow shock's evolution; Thirdly, the CME-driven shock has an
extended plane shock front structure near the Earth, but the
Earth-bow shock front has a stationary bow shock geometry. We
predict those differences would produce a different non-linear
properties including the evolution of the shock fine structure,
energy injection rate and even the energy spectral shape. This paper
will focus on the understanding some of the non-linear properties of
the planetary CME-driven shock. The 14-Dec-2006 shock event was
fortuitous as it provides us an opportunity for applying the
dynamical Monte Carlo package code, which was developed on the
Matlab platform \citep{wy11}.

This paper is structured as follows: In Section \ref{Observations},
we present the specific observations for the 14-Dec-2006 CME-driven
shock event; The detailed description of the method is given in
Section \ref{The Model}; We present the simulated results and
discussions in Section \ref{Results}; Finally, Section
\ref{sec-summary} presents the summary and some conclusions.


\section{Observations}
\label{Observations}

\begin{figure}[h]\center
{\includegraphics[width=3.0in,angle=0]{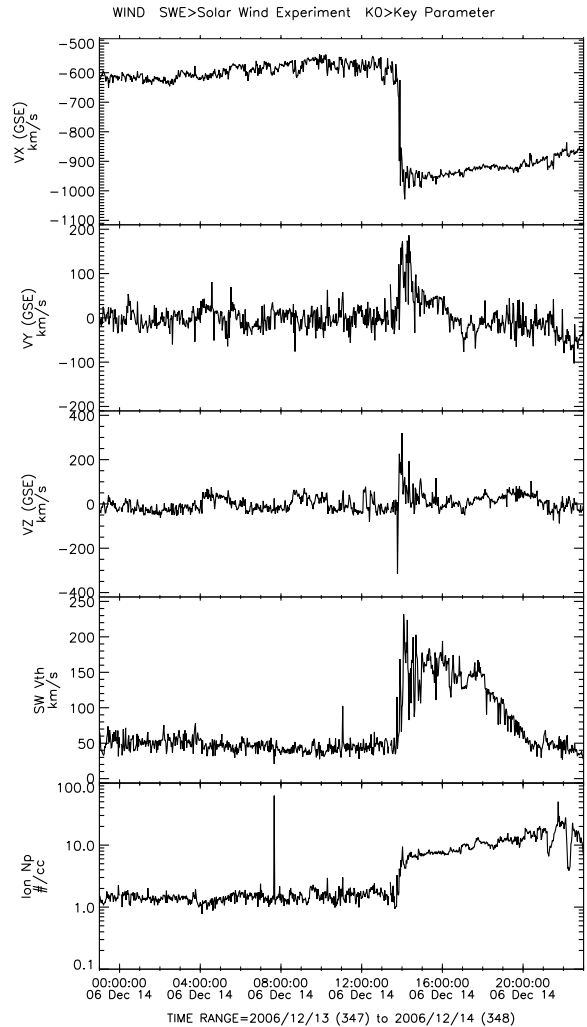}} \caption{The
plot shows the key parameters of the 14-Dec-06 shock event in the
Wind spacecraft, and the data come from $ http
://cdaweb.gsfc.nasa.gov/cdaweb$.} \label{fig:wind}
\end{figure}

\begin{figure}[h]\center
{\includegraphics[width=3.5in,angle=0]{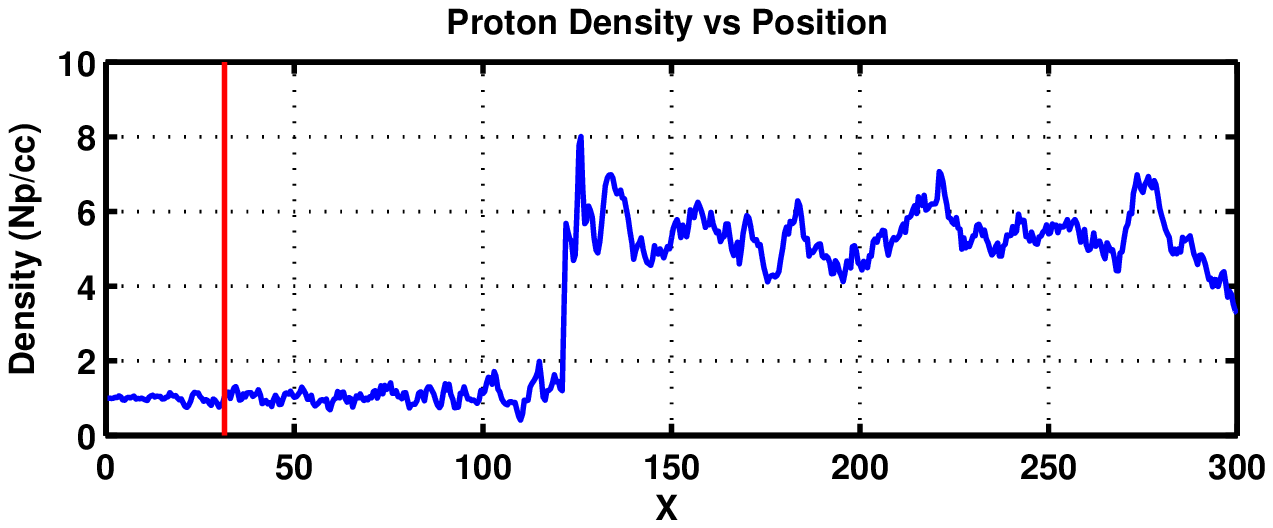}}\vspace{-1.3in}
{\includegraphics[width=3.5in,angle=0]{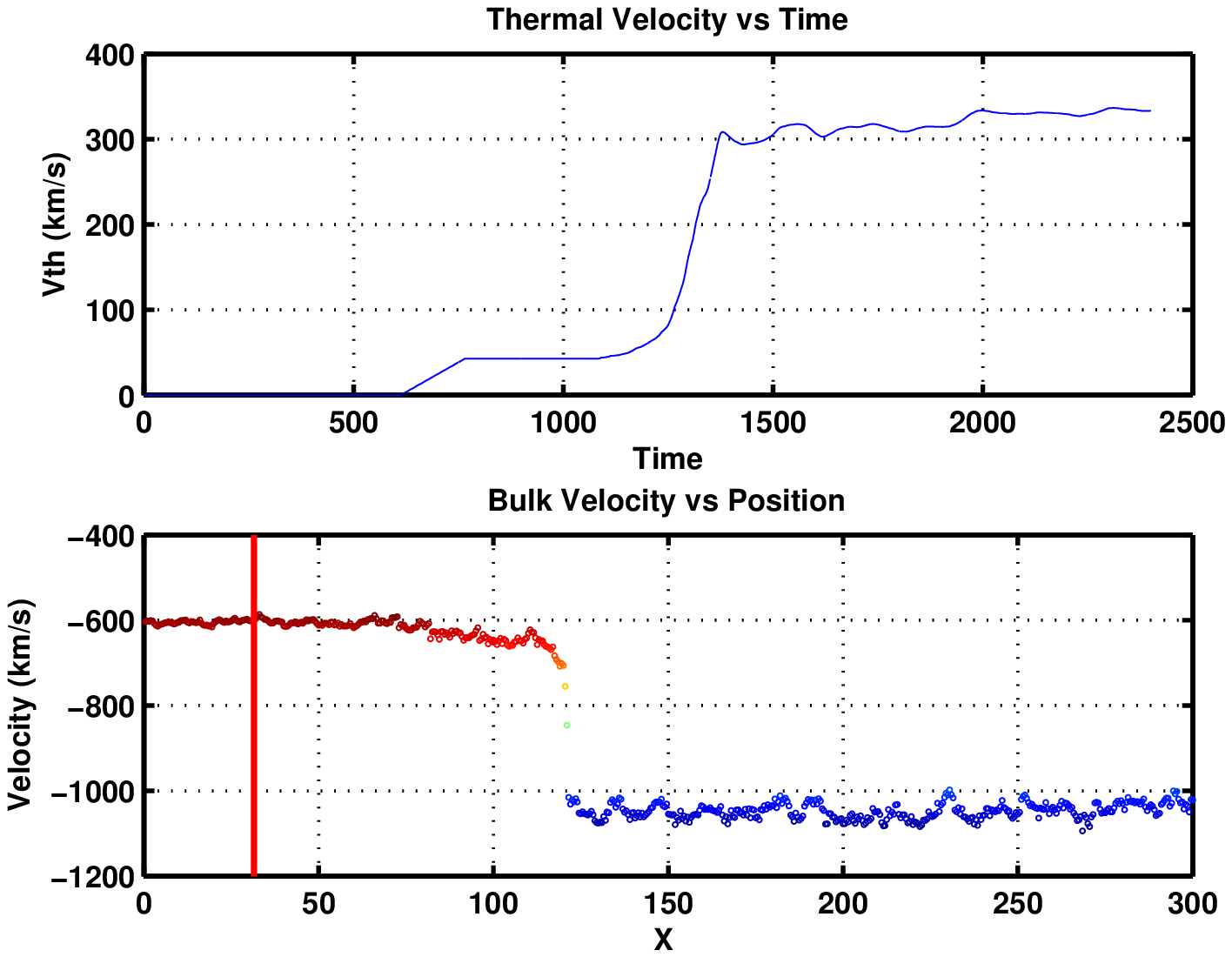}}
\caption{The upper panel represents the proton density profile vs
its position at the end of the simulation. The middle panel denotes
the solar wind thermal velocity profile in the local frame vs the
time. The lower panel indicates the bulk flow speed profile vs its
position at the end of the simulation. The vertical lines in the
upper and lower panels both show the final FEB position at the end
of simulation.} \label{fig:vth}
\end{figure}

The unusual group of CME-driven shock events of solar cycle 23 was
observed in December 2006 at the solar active region 10930. Halo
CMEs were observed by the LASCO coronagraphs in association with the
events of 13 and 14 December, with speeds of 1774km/s and 1042km/s,
respectively. Because the 14 December solar event was better
magnetically connected to the Earth, so it provided the  best
opportunity for testing the nonlinear effect and efficiency of the
diffusive shock acceleration (DSA) mechanism. As shown in Figure
\ref{fig:wind}, an overview of key parameter observations from the
Proton Monitor (PM) instruments on Wind/SWE for the CME shock event
of 14 December 2006 is given in detail. This event originated on the
western hemisphere of the Sun. It showed an abrupt fluctuation in
intensity of proton density and solar wind thermal speed during the
decay of the 13 December solar event. The initial particle increase
following the 14 December solar event was seen in the higher energy
range, as expected for velocity dispersion. There was also a higher
background at the lower energy associated with the 13 December solar
event and the related shock \citep{von09}. Simultaneously, many
spikes were also detected to be superposed on the radio continuum in
the frequency range 2.6-3.8 GHz by the digital spectrometers of
NAOC. These spikes were found to have complex structures associated
with other radio burst signatures connecting with the in-situ SEP
event observations \citep{wsj2008}.

Both Wind and ACE were in orbit around the {\it L1} Lagrangian point
$\sim$ 1.5 million km upstream of the Earth. Similar intensity
modulations were observed at Wind and ACE. As has also been noted by
\cite{mbm08}, the variations of the particle intensity and smooth
magnetic fields observed by near-Earth spacecraft occurred in the
duration of the interplanetary coronal mass ejection (ICME) driving
the shock on 14 December which was related to the 13 December solar
event. Solar wind observations from ACE show evidence of the
presence of the ICME, which had the enhanced magnetic field and a
smooth rotated  ``magnetic cloud" in the upstream shock and the
intervening sheath region, respectively. According to the Wind
magnetic cloud list, the axis orientation of this ``magnetic cloud"
was $\theta$ = $27^{o}$, $\phi$ = $85^{o}$. In addition,
\cite{liu08} estimate that the ``cloud" axis direction was $\theta$
=$-57^{o}$, $\phi$ = $81^{o}$ in GSE coordinates. Thus, both agree
that the axis was closely aligned west to east but differ in whether
it was inclined north or south, most likely because different
intervals were considered in their analysis.

\begin{table}
\begin{center}
  \caption{The Parameters of the Simulated Shock\label{tab:para}}
  \begin{tabular}{|c|l|c|c|}
  \hline
     \multicolumn{2}{|c|}{Simulated Parameters} & Dimensionless value & Scaled value
     \\  \hline
    \multirow{6}{*}{Physical }   & Upstream bulk speed & $u_{u}$=-0.4467 & -600km/s  \\
    &Downstream bulk speed & $u_{d}$=-0.7742 & -1042km/s \\
    &Relative inflow velocity & $\Delta u$=0.3275& 442km/s \\
    &Inflow sonic Mach number & M=17.5 & ... \\
    &Thermal speed & $\upsilon_{0}$=0.0342 & 46km/s \\
    &Scattering time & $\tau$=0.3333 & 0.052s  \\
\hline    &Box size & $X_{max}$=300 & $10R_{e} $ \\
\multirow{6}{*}{Numerical }     &Total time & $T_{max}$=2400 & 6.3minutes  \\
    &Time step size & $dt$=1/30 & 0.0053s  \\
    &Number of zones & $nx$=600 & . . . \\
    &Initial particles per cell & $n_{0}$=650 & . . .  \\
    &FEB distance & $X_{feb}$=90 & $3R_{e}$ \\
\hline
\end{tabular}
 \end{center}
 {Notes: The physical parameters are taken from the Wind spacecraft, and the
 numerical parameters are decided by the 14-Dec-2006 CME-driven shock.}

\end{table}

\section{The method}
\label{The Model}
\subsection{Physical model}

\begin{figure}\center
{\includegraphics[width=4.5in]{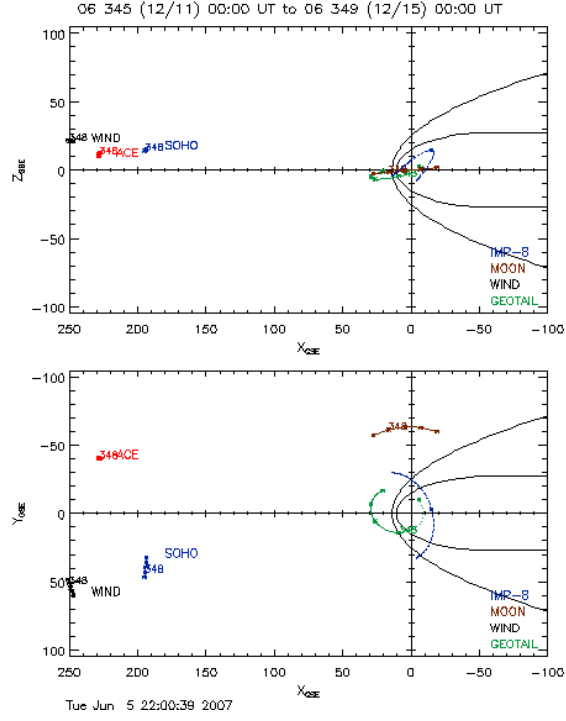}} \caption{The
diagrams show the realistic obits of the near-Earth spacecraft. The
obit data are taken from $ http ://cdaweb.gsfc.nasa.gov/cdaweb$. }
\label{fig:orbit}
\end{figure}

We consider a plane-parallel shock where the supersonic flow moves
from the Sun to the Earth (in the rest frame) along the
\textit{x}-axis direction. The shock was observed by Wind, SOHO, and
ACE spacecraft near the Earth in the location of the first
Lagrangian point $L1\sim$ 1.5 million km ($\sim$ 250$R_{e}$, where
$R_{e}$ is the radius of the Earth) upstream of the Earth on 14
December. All trajectories of the spacecraft in the 348th day
corresponding to the 14 December 2006 are shown in Figure
\ref{fig:orbit}. With the CME-driven shock propagating from the Sun
along \textit{x}-axis to the Earth, its shock front were encountered
by  Wind, SOHO, and ACE spacecraft located in $X_{GSE}$ between the
250$R_{e}$ and 180$R_{e}$ upstream to the Earth.  These three
spacecraft moved about 10$R_{e}$ distance in their obits on the
348th day. The distances of all these three spacecraft from the
Sun-Earth line were within 50Re  along the $Y_{GSE}$ and $Z_{GSE}$
directions. The 14-Dec-2006 shock event originated from the western
hemisphere of the Sun with an interplanetary `` magnetic cloud" axis
orientation of $\theta$ = $27^{o}$, $\phi$ = $85^{o}$. And the
actual trajectory of  Wind spacecraft at that moment is just tangent
to the Sun-Earth line with an angle $\phi$ = $80^{o}$ as shown in
the lower panel of the Figure \ref{fig:orbit}. As far as the
position of  Wind spacecraft is concerned, the observed CME-driven
shock is just a parallel diffusive shock.  So the observation of
Wind spacecraft provided an example of semi-parallel shock for
applying our dynamical Monte Carlo code to understand the particle
injection problem of DSA theory.

The important physical parameters of this simulation include the
upstream bulk flow velocity ($u_{u}$), the downstream bulk flow
velocity ($u_{d}$), the relative bulk inflow velocity difference
($\Delta u$), the inflow sonic Mach number ($|u_{d}|/c_{s}$), which
is 17.5 (where $c_{s}\equiv (\gamma kT/m)^{1/2}$, $c_{s}$ is the
upstream sound speed), the upstream thermal velocity [$v_{0}\equiv
(kT/m)^{1/2}$], and the constant scattering time ($\tau$), which is
2/5 times of the scattering time ($\tau_{0}=0.13s$) used by {Knerr,
Jokipii \& Ellison (1996)} in the Earth-bow shock simulation. Since
there are some differences in the shock geometry between the
CME-driven shock (i.e. at {\it L1} point) and the Earth-bow shock,
we chose the scattering time 0.4 times smaller than that in the
Earth-bow shock, which is equivalent to a 2.5 times larger FEB
distance than that in the Earth-bow shock. The specific physical
parameters and numerical parameters are listed in the Table
\ref{tab:para} with their dimensionless values and scaled values,
respectively.

\subsection{Mathematical model}
\begin{figure}\center
{\includegraphics[width=2.5in,angle=-90]{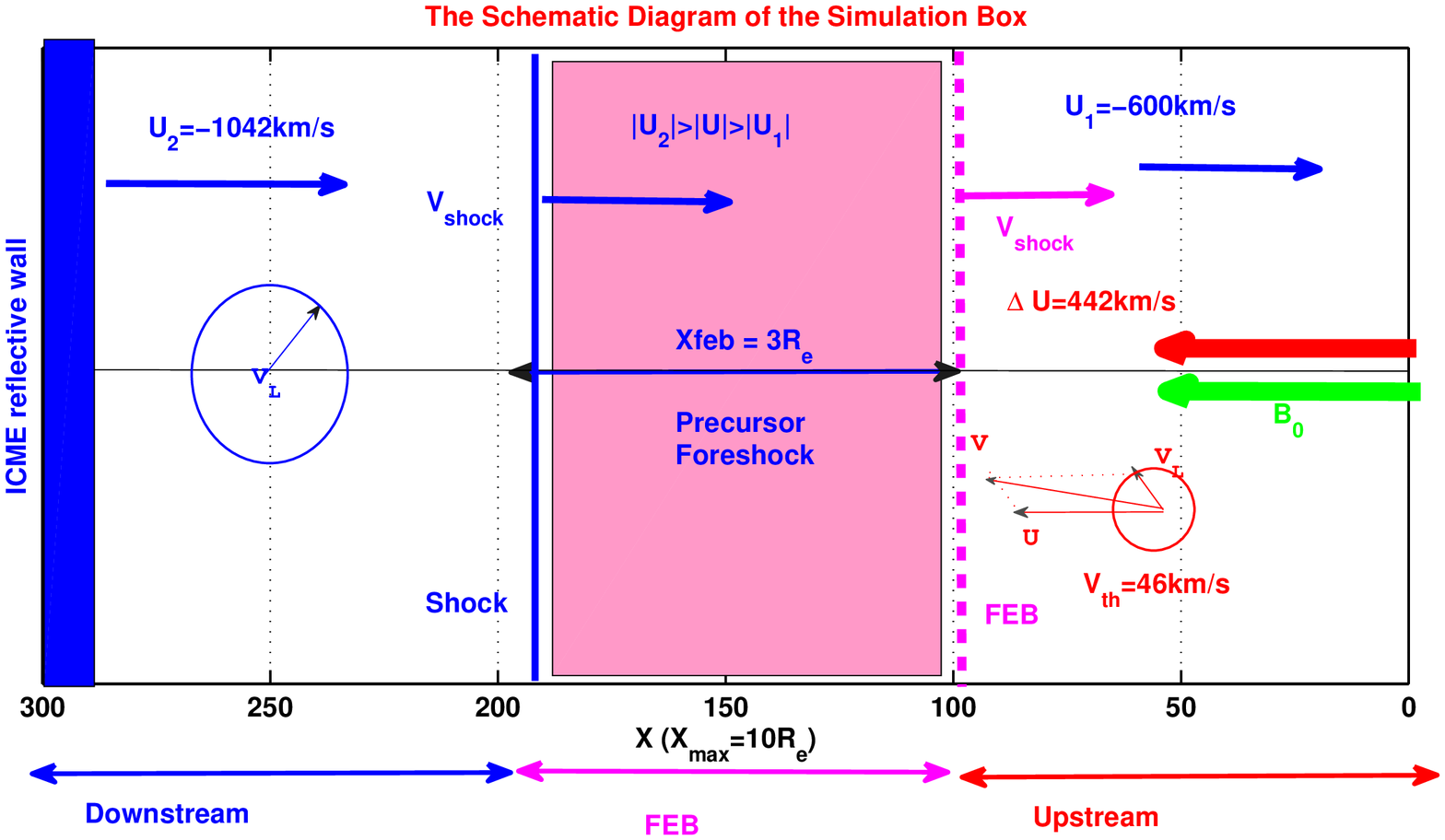}}
\caption{The schematic diagram of the simulation box. The  left
reflective wall acts ICME  to produce the ICME-driven shock
prorogating from the left boundary to the right boundary of the
box.} \label{fig:schematic}
\end{figure}

According to the observed 14-Dec-2006 CME-driven shock, the
schematic diagram of the simulation box can be designed as
one-dimensional parallel shock along the x-axis direction. As shown
in Figure \ref{fig:schematic}, the initial particles with a relative
bulk flow speed difference ($\Delta u$) move from the right to the
left. The initial particles have a background Maxwellian thermal
distribution with an initial temperature ($T_{0}\equiv
mv_{0}^{2}/k$) in the local frame. To begin and maintain the shock
simulation, particles are assumed to flow into the simulation box
from the pre-inflow box (PIB) at the right boundary. Then, with the
continuous particle flow moving forward one time step, only those
particles which move into the main simulation box are actually added
to the simulation. This process naturally leads to a flux-weighted
inflow population. At the left boundary of the box, a reflective
wall acts to produce a CME-driven shock moving from the left to the
right. Considering the geometry of the 14 December shock event, we
just follow the parallel component of the CME-driven shock observed
in  Wind spacecraft.

Figure \ref{fig:schematic} also shows one typical particle and its
local ($V_{L}$) or box frame ($V$) velocity in the upstream region
and downstream region, respectively. The majority of the incoming
particles cross the shock front only once from the upstream region
to the downstream region and stay in the downstream region. A small
portion of the particles can effectively scatter off the resonant
MHD wave self-generated by the energetic particles and return to the
upstream region to re-cross the shock front for additional energy
gains \citep{LPM2004}. Thus an anisotropic energetic particle
distribution, but not a strict Maxwellian distribution, is produced
in the diffusive regions. It is this elastic interaction between
individual particles and the collective background that allows the
Fermi acceleration to occur.

The position of the FEB could coincide with a location upstream of
the shock where particles are no longer able to scatter effectively
and return to the shock.  A reasonable FEB farther out in front of
the shock  moves, companying with the shock front, at the same shock
evolutional velocity $V_{sh}$. This constant FEB distance is acted
to inform a precursor region which is showed by the shadow area in
the middle of the Figure \ref{fig:schematic}.  If one particle
archives to the highest energy, and exceeds the position of the FEB
in front of the shock, it will be taken as the escaped particle and
removed from the simulation system.  According to the actual motion
of  Wind spacecraft in the duration, the spacecraft moved about
10$R_{e}$ distance in their obits on the 348th day. To simulate the
shock formation and evolution, the total length of the simulation
box is set to be 10$R_{e}$, the length of the FEB is set to be
$\sim$3 $R_{e}$. The scattering time is set to be $\tau=0.052s$ on
the basis of the Earth-bow shock model \citep{kje96}.

The important numerical parameters include the box size ($x_{max}$),
the time to evolve the whole system ($t_{max}$), the number of grid
zones ($n_{x}$), the initial number of particles per zone ($n_{0}$),
and the size of the time step ($dt$). Because of the similar
character of the plasma flow near the Earth, we take some numerical
parameters as in the Earth-bow shock model \citep{kje96}.
Specifically, the total box length $x_{max}=300$ is divided into
$n_{x}=600$ grids, with each grid length being $\Delta x=1/2$; the
total time $t_{max}=2400$ is divided into $n_{t}=72000$ steps by
$dt$, with each step being $dt=1/30$. All numerical parameters are
listed in Table \ref{tab:para}. The physical parameters and  the
numerical parameters constitute the whole simulation parameter list.
As shown in Table \ref{tab:para}, each dimensionless value  is
corresponding to its scaled value. The scale factors for distance,
velocity, and time are $x_{scale}=10R_{e}/300$, $v_{scale}=442km
s^{-1}/0.3275$, and $t_{scale}=x_{scale}/v_{scale}$, respectively.

The presented simulations apply the same steps like the Earth-bow
shock model \citep{kje96} including three sub steps: (i) All the
particles moving with their velocities in the simulation box along
the $x$ axis direction. (ii) Summing particle masses and velocities
over each background computational grid. (iii) Invoking the
scattering angular distribution law. The particle diffusive
processes in the presented simulations are dominated by the Gaussian
scattering angular distributions. The scattering rate is
$R_{s}=dt/\tau$, which implies that only this fraction of particles
is able to scatter off the scattering center frozen in the
background fluid.  The candidate does not change its route until it
is selected to scatter once again. So the particle's mean free path
is proportional to the local thermal velocities in the local frame
with
\begin{equation}
\lambda = V_{L}\cdot \tau \label{eq_step3b}.
\end{equation}
For an individual proton, the grid-based scattering center can be
seen as a sum of individual momenta. So these scattering processes
can be taken as the elastic collisions. In an increment of time,
once all of the candidates complete these elastic collisions, the
momentum of the grid-based scattering center is changed.  In turn,
the momentum of the grid-based scattering center will affect the
momenta of the individual particles in their corresponding grid in
the next increment time. One complete time step consists of the
above three substeps. The total simulation temporally evolves
forward by repeating this time step sequence. To calculate the
scattering processes accurately and produce an exponential mean free
path distribution, the time step should be less than the scattering
time (i.e. $dt<\tau$).

The scattering angles consist of two variables: $\delta\theta$ and
$\delta\phi$. Once a particle has a collision with the background
scattering centers, its pitch angle becomes
$\theta'$=$\theta$+$\delta\theta$, and the azimuthal angle becomes
$\phi'$=$\phi$+$\delta\phi$, where $\delta\theta$ is the variation
in the pitch angle $\theta$, and $\delta\phi$ is the variation in
the azimuthal angle $\phi$. The pitch angles $\theta$ and $\theta'$
are both in the range $0\leq\theta,\theta'\leq \pi$, and azimuthal
angles $\phi$ and $\phi'$ are both in the range $0\leq\phi,\phi'\leq
2 \pi$ on the unit sphere. The variation in the pitch angle
$\delta\theta$ and azimuthal angle $\delta\phi$ are composed of the
scattering angle, and its anisotropic character is described by the
Gaussian function $f(\delta\theta,\delta\phi)$. Here, we will just
present the results of the CME-driven shock using the Gaussian
scattering angular distribution with a standard deviation value of
$\sigma=\pi$.

\section{Results}\label{Results}
\subsection{Data analysis}\label{subsec:eng}
Since the individual particle energy can be examined at any given
time in the simulation, so the energy function over time can be
obtained.  At first, we can calculate the necessary energy functions
for further analysis. In this simulation, we obtained the total
energy function in the box, the loss energy function escaped from
the FEB, and the injected energy function which is the energy
summation of the injected energetic particles from the downstream
region at the local velocity of $V_{L}=U_{0}$ over time. Then, at
the end of the simulation, we obtained the final values of the total
energy $E_{tot}=3.5666$, the energy loss $E_{loss}=0.2010$, and the
energy injection $E_{inj}=0.5464$, respectively. The final energy
injection rate $R_{inj}$, which represents the acceleration
efficiency, can be defined by the formula as follows.
\begin{equation}\label{eq:Rinj}
    R_{inj}=E_{inj}/E_{tot}
\end{equation}
The injection rate is so important for a CME-driven shock, because
it is connected with the facts that the shock how distribute itself
energy to accelerate cosmic ray (CR) and to ``heat" the thermal
background plasma. By a series of simulations, we give a plausible
injection rate with a value of $R_{inj}=15.32\%$ for the 14-Dec-2006
shock. Under this condition, we obtain the maximum energy particle
with the dimensionless value of $VL_{max}=20.2609$ and the scaled
value of $E_{max}=3.8684MeV$. In addition, because there exists some
energy losses in the simulation system, the shock fine structures do
not completely agree with the situ observations. Finally, according
to the DSA theory, the energy spectrum index can be calculated based
on the simulated compression ratio (i.e., $\Gamma=(r+2)/[2(r-1)]$).
We calculated the total energy spectral index with a value of
$\Gamma_{tot}=0.8406$ and the vicious subshock energy spectral index
with a value of $\Gamma_{sub}=1.1074$, respectively.

\begin{figure}\center
 \includegraphics[width=3.5in]{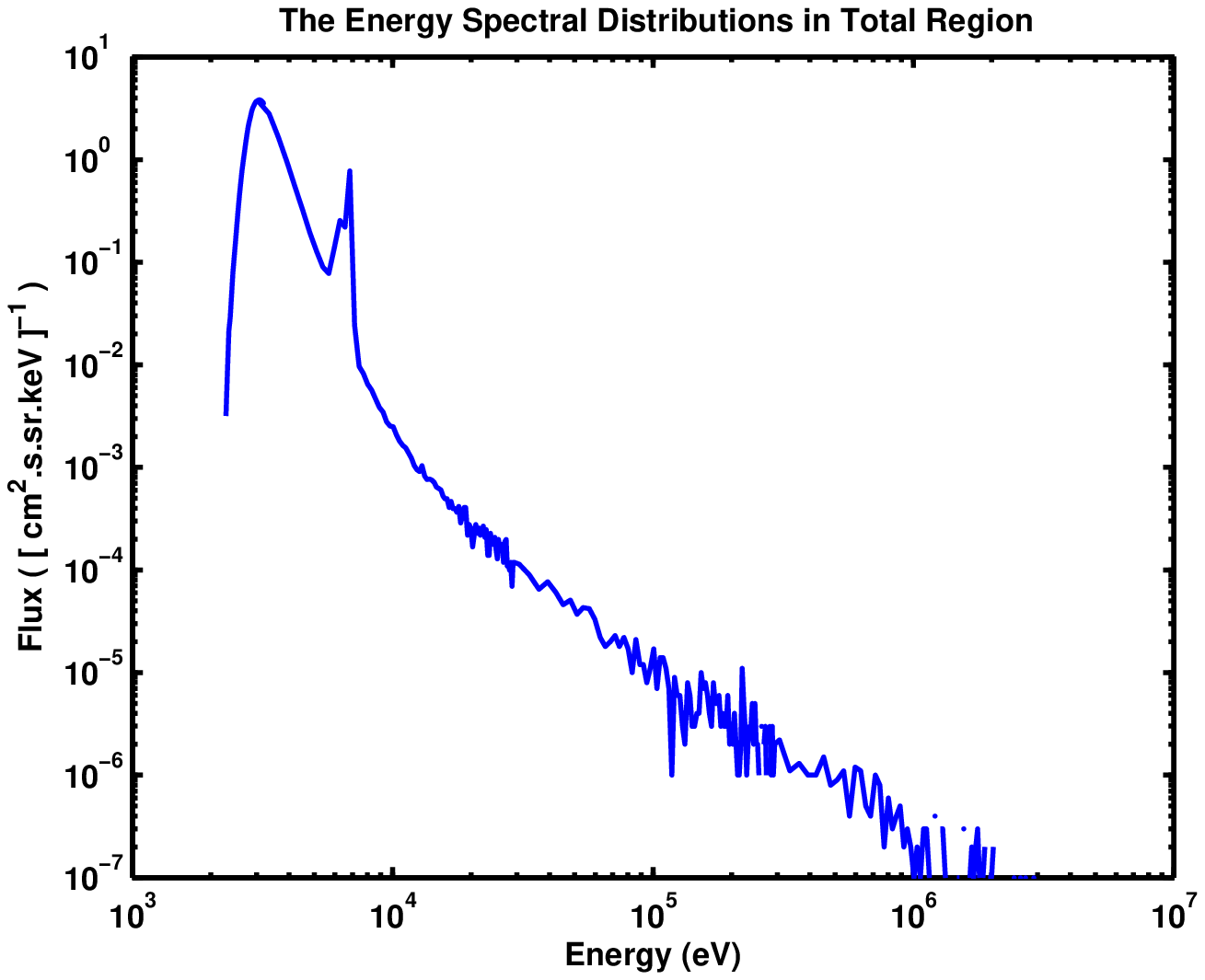}
  \includegraphics[width=3.5in]{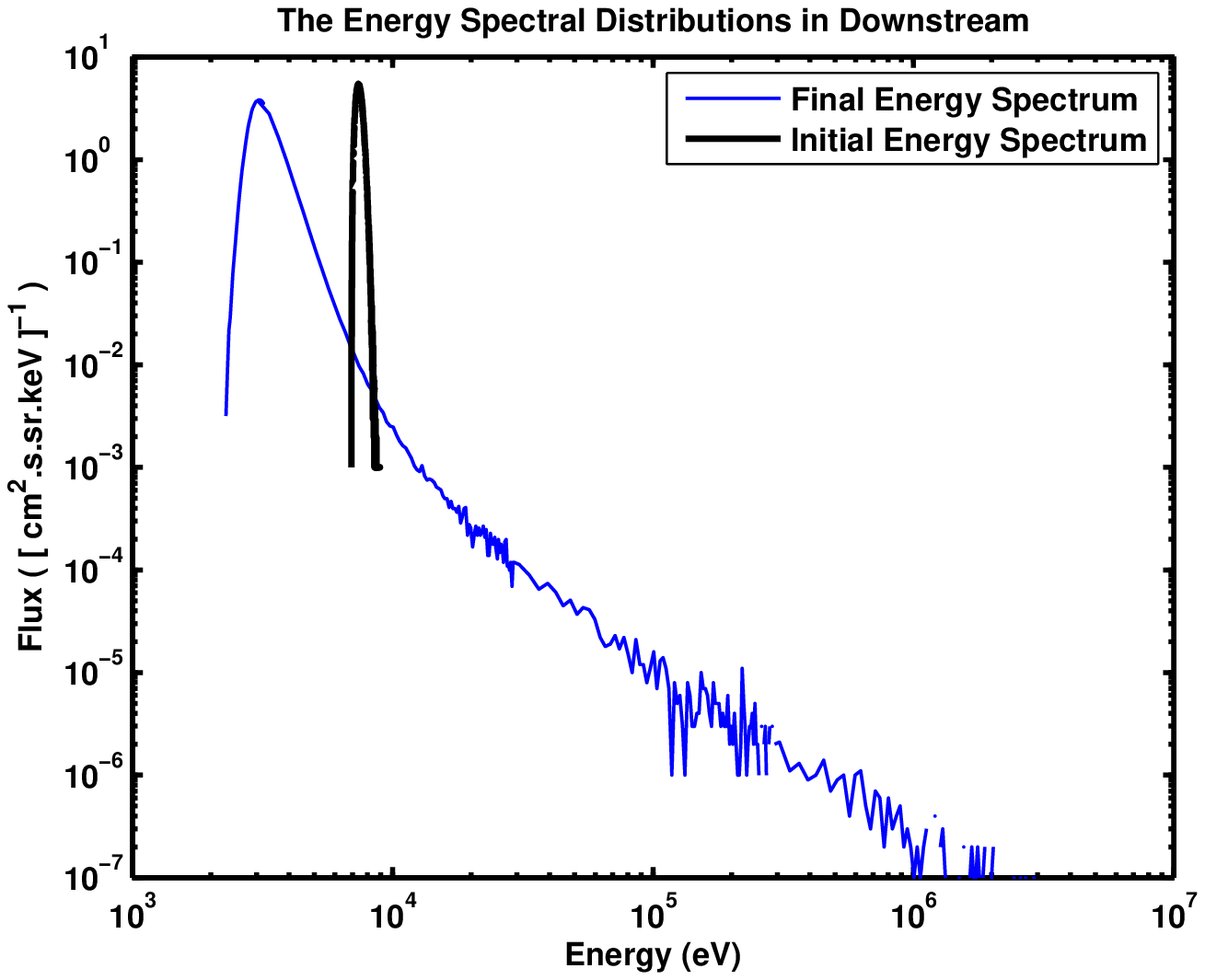}
\caption{The extend energy spectra in two plots are calculated over
the entire simulation region and the only downstream region at the
end of the simulation, respectively. The solid extended curve with a
``power-law" tail in each plot represents the final shocked energy
spectrum. The thick curve with a narrow peak denotes the initial
Maxwellian energy spectrum.}\label{fig:spec}
\end{figure}
\begin{figure}\center
{\includegraphics[width=4.0in,angle=0]{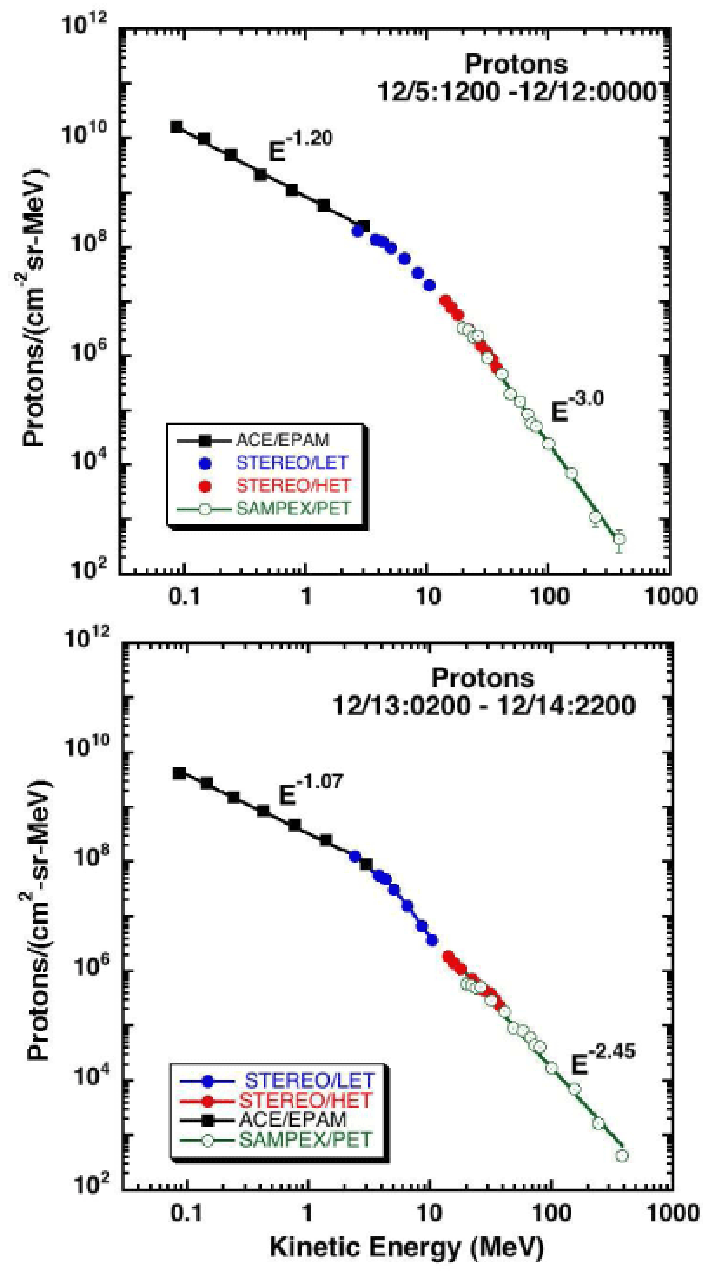}}
\caption{Fluency spectra of the protons measured in the two largest
December 2006 SEP events by multiple spacecraft. The energy range is
from 5 to 100 MeV adapted from  Mewaldt et al. (2008). }
\label{fig:sep}
\end{figure}

Figure \ref{fig:spec} shows the simulated energy spectra. The first
plot shows the energy spectrum with the ``double-peak" structures
averaged over the entire simulation box at the end of the
simulation. The second plot shows an energy spectrum with a
``power-law" tail averaged only over the downstream region at the
end of the simulation. In this plot, the thick solid curve with a
narrow peak represents the initial Maxwellian thermal energy
spectrum in the shock frame. As viewed from the first plot, the
double peaks imply that there exist two thermal particle
distributions in the entire simulation box: the left peak represents
the ``heated" downstream flow distribution and the right peak
represents the Maxwellian distribution in the unshocked upstream
flow.  Turn to look at the second plot, we find the final extend
energy spectrum at the left of the panel shows several decade times
wider than that in the initial energy spectrum at the right of the
panel. This means that there exist a large temperature difference
between  the shocked downstream region and the unshocked upstream
region.

As shown in Figure \ref{fig:sep}, the spectra of protons in the two
largest December 2006 SEP events by ACE, STEROEO, and SAMPEX
instruments are reported. The particle intensities started
increasing at the beginning of the December 5 and the other at the
onset of the December 13 event.  These two events both have spectra
that roll over in a similar fashion beyond $\sim$50MeV, as in the
20-Jan-2005 SEP event \citep{mewaldt08,WZZ2009,Bartoli2012,WYF2010}.
The fitting energy spectral shape of the 12-14 December 2006 events
are showed in the lower panel and the spectral index is marked  as a
value of $E^{-1.07}$ in the lower energy range. The predicted
subshock energy spectral index ($\Gamma_{sub}=1.1074$) from our
simulation is consistent with the observed energy spectral index in
the lower energy range.  Owing to computer constraints on the size
of the simulation grid, this simulated energy spectrum is just in
the range from keV to MeV. We speculate that the second ``roll over"
on the higher energy spectrum could be obtained if a larger
simulation box size is used. This will be investigated in a future
simulation. There are two conditions suggesting that the
``roll-over" would be reproduced at high energy: (1) The FEB
distance decides the maximum diffusive length (i.e.
$FEB\equiv\lambda_{max}=\tau\cdot p_{max}$). If we enlarge the FEB
distance and the total simulation box, we can obtain the larger
$P_{max}$ in the new simulation system. (2) In the Figure
\ref{fig:sep}, we can see the first power law $E^{-1.07}$ as the
input function of the second power law $E^{-2.45}$ at the high
energy range. Simultaneously, in the Figure \ref{fig:spec}, we can
see the heated Maxwellian thermal distribution, which would be
represented by a similar power law $E^{-0.5}$ averaged over the
respective energy range, as the input function of the first power
law $E^{-1.1074}$.

\subsection{Shock structures}
At the end of the simulation, the simulated shock with the specific
parameter values are given as follows: the shock position
$X_{sh}=121.5$, the FEB position $X_{f}=31.5$, the shock evolutional
velocity $V_{sh}=-0.0744$, the subshock velocity $V_{sub}=0.2103$,
total compression ratio $r_{tot}=5.4034$, subshock's compression
ratio $r_{sub}=3.4697$, total energy spectral index
$\Gamma_{tot}=0.8406$,  subshock's energy spectral index
$\Gamma_{sub}=1.1074$, particle injection rate $R_{inj}=15.32\%$,
energy loss $E_{loss}=0.2010$, the maximum energy particle's local
velocity $VL_{max}=20.2609$, and the maximum energy particle
$E_{max}=3.8684MeV$.

\begin{figure}\center
       \includegraphics[width=3.5in, angle=0]{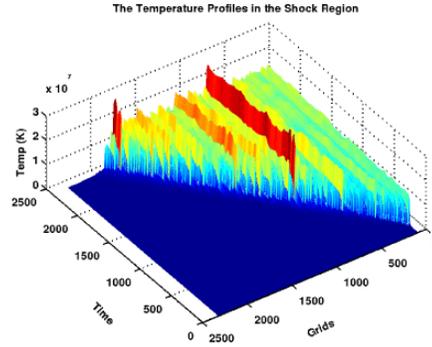}
\caption{The mesh plot represents the evolutional bulk flow
temperature profiles in their positions with the time. The lower
temperature represents the upstream bulk flow. The higher
temperature represents the downstream bulk flow. The apparently
distinguished boundary traces the shock front positions with
time.}\label{fig:shock}
\end{figure}

First, we present the entire shock evolution with the temperature
profile of the time sequences as shown in Figure \ref{fig:shock}.
The supersonic continuous inflow with an initial Maxwellian thermal
velocity $v_{0}$ in each grid evolves from the begin to the end of
the simulation. Their kinetic energies are translated into the
random thermal energetic particles by the ``heating" processes in
the downstream region resulting in a distinct enhancement
temperature profiles in the shock front with the time. The profile
of the thermal temperature shows the upstream averaged temperature
of $T_{0}=2.5\times10^{5}K$ and the downstream averaged temperature
of $T_{d}=9.0\times10^{6}K$. This means that the CME-driven shock
can ``heat" the background plasma efficiently and provide the
first-order Fermi acceleration mechanism  by crossing the shock
front for accelerating the particles, which are injected from the
``heated" downstream region into the precursor region.

\begin{figure}\center
       \includegraphics[width=3.5in, angle=0]{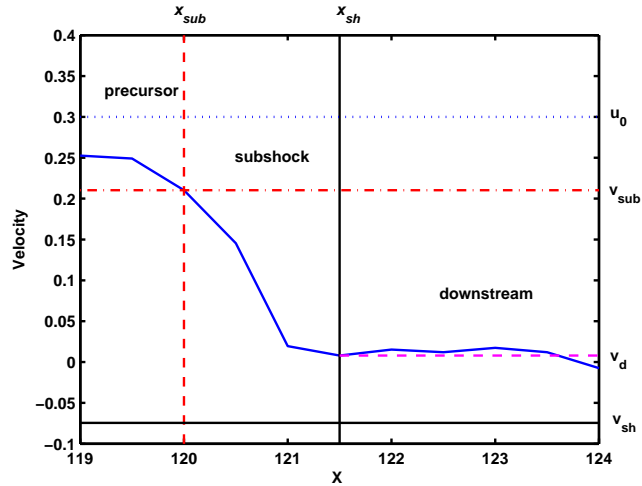}
\caption{The bulk flow speed fine structure of the simulated shock
at the end of the simulation. The vertical dash line and vertical
solid line split the entire region into three sections: precursor,
subshock and downstream region. }\label{fig:subshock}
\end{figure}

Figure \ref{fig:vth} shows a group of profiles of the physical
parameters in the simulation. From top to bottom, the upper panel
shows the proton density profile vs its position. The proton density
is presented by the scaled value.  The enhanced density flux
apparently appears in the position of the shock front. The intensity
of the density in the downstream is about five times larger than
those in the unshocked upstream bulk flow. By comparing the proton
density profiles between the the Figure \ref{fig:vth} and the Figure
\ref{fig:wind}, the simulated bulk flow has a lightly higher proton
density intensity in the downstream bulk flow than that in the
observed downstream bulk flow. The middle panel in the Figure
\ref{fig:vth} denotes that the thermal velocity profile evolutes
with the time. The profile at time of T$\sim$600 (it is zero before
the simulation time T$<$600, take account of the injection from the
PIB) has an initial Maxwellian thermal velocity of $v_{0}=46km/s$
until it is shocked. After the profile is shocked as shown in middle
panel of the Figure \ref{fig:vth}, it reaches an average thermal
velocity with a value around $<v_{d}>=300km/s$ till the end of the
simulation. Also the thermal velocity profile shows a slightly
larger enhancement than that from the observation by  Wind
spacecraft. Although the simulated proton density and the solar wind
thermal velocity are slightly larger than those from in-situ
observations, we suggest that this is caused by the insufficiency of
particles in the simulation.  We have demonstrated that it is the
case using a series of simulations with different initial number of
particles per cell. The lower panel indicates the profiles of the
bulk flow speed vs its position at the end of the simulation. The
profile shows an upstream bulk flow speed $U_{u}=-600km/s$ and the
downstream bulk flow speed $U_{d}=-1042km/s$ which are followed by
the observations. The complex shock front fine structure will be
showed in Figure \ref{fig:subshock} at the end of the simulation.
The final evolutional positions of the FEB and the shock front are
$X_{f}=31.5$ and $X_{sh}=121.5$ in the x-axis, respectively. The
distance between these two locations is just the size of the
precursor region where the particle acceleration processes occur. It
is just this region slowed the incoming upstream bulk flow speed
$U_{u}$ down to the downstream bulk flow speed $U_{d}$. The bulk
flow speed in precursor region is between the two bulk flow speeds
(i.e. $U_{u}>U_{p}>U_{d}$). From our simulation, we can see that the
particle acceleration process and the ``back pressure" due to the
energetic particles occurred mostly in precursor region which
results in a non-linear shock structure that is characterized by a
bulk flow speed gradient. According to the evolutional shock front
position $X_{sh}$ with the time, we can calculate the shock
evolutional velocity $V_{sh}$ as follows.
\begin{equation}\label{eq:vsh}
V_{sh}=\frac{|X_{max}-X_{sh}|}{T_{max}},
\end{equation}
where, the $X_{max}$ is the total length of the simulation box, and
the $T_{max}$ is the total simulation time. Then, we are able to
calculate the total shock compression ratio in the shock frame as
follows.
\begin{equation}\label{eq:rtot}
    r_{tot}=\frac{\Delta U+|V_{sh}|}{|V_{sh}|},
\end{equation}
where $\Delta$U is the relative bulk flow speed between the upstream
and downstream, $V_{sh}$ is the shock velocity.

Figure \ref{fig:subshock} shows the shock fine structure with the
bulk flow speed near the shock front at the end of the simulation.
$V_{sub}=0.2103$ shows the bulk flow speed of the subshock,
$V_{d}\simeq 0$ shows the bulk flow speed of the downstream region,
$V_{sh}=-0.0744$ represents the value of the opposite shock
evolutional velocity, and $U_{0}=0.3275$ represents the incoming
bulk flow speed with a related bulk flow speed difference of $\Delta
U$. All zones of the precursor, subshock and downstream are divided
by a vertical dashed line and a solid line in the plot. These three
zones constitute the total shock fine structure in the simulated
shock region.  The smooth precursor has a long scale in the range
from the subshock's position $X_{sub}$ to the FEB position $X_{f}$,
which is invisible, beyond the left boundary of the plot. This zone
is called the diffusive zone where the bulk flow speed will be
slowed by the ``back pressure" of the accelerated particles. The
subshock region with a narrow scale of a three-grid-length has a
deep drop of the bulk flow speed, in which the bulk flow speed vary
from the subshock velocity $V_{sub}$  to the downstream velocity
$V_{d}$. The scale of the three-grid-length is almost identical to
the averaged thermal mean free path over the downstream region. The
subshock velocity $V_{sub}$ is decided by the horizontal dot-dashed
line with a value of $V_{sub}=0.2103$. The downstream velocity
$V_{d}$ is marked with a horizontal dashed line at the end of the
simulation, which should be with an averaged value of $<V_{d}>=0$
over the entire simulation time in the box frame. The nagative shock
evolutional velocity marked with a horizontal solid line shows an
value of $V_{sh}=-0.0744$. We can calculate the subshock's
compression ratio according to Rankie-Hongniout relationships in the
shock frame as follows.
\begin{equation}\label{eq:rsub}
    r_{sub}=\frac{V_{sub}+V_{sh}}{<V_{d}>+V_{sh}},
\end{equation}
where, we take the averaged value of the downstream velocity $<
V_{d}>$ equal to zero.

\section{Summary and conclusions}\label{sec-summary}
In summary, we performed a dynamical Monte Carlo simulations on the
14-Dec-2006 CME-driven shock using an anisotropic scattering law.
The specific temperature profile, shock fine structures, particle
injection function, as functions of time, are presented. We examined
the correlation between the energy injection and the shock energy
translation processes of the interplanetary CME-driven shock.
Simultaneously, we find the simulated CME-driven shock  energy
spectrum provides a good fit to the observations from the multiple
spacecraft.

In conclusion, the dynamical Monte Carlo simulation of the
14-Dec-2006 CME-driven shock demonstrates that the energy spectrum
is affected by the specific non-linear factor of the DSA. This paper
focus on the energy injection, which is one of important nonlinear
effects of the DSA. By calculating the energy injection rate of the
CME-driven shock, we can understand how the CME-driven shock
distributes its shock energy to accelerate the energetic particles
by first-order Fermi acceleration mechanism as well as how it heats
the solar wind background bulk flow at a certain efficiency. We give
an energy injection rate of $R_{inj}=15.32\%$ in the 14-Dec-2006
CME-driven shock. We guess that this predicted injection rate could
satisfy the required energies of the observed SEP events, which
should be released from the CME-driven shock.

\begin{acknowledgements}
The work was supported by the Chinese Academy Sciences NSFC grant
10921303, and the National Basic Research Program of the MOST Grant
(No. 2011CB 811401).
\end{acknowledgements}

\end{document}